\documentclass[preprint,showpacs,preprintnumbers,amsmath,amssymb, superscriptaddress,longbibliography,nofootinbib]{revtex4-1}
\usepackage{graphicx} 
\usepackage{xcolor}
\usepackage{amsmath}
 \usepackage{hyperref}
\usepackage[top = 2cm, bottom = 2cm, right = 2cm, left =2cm]{geometry}

\begin{document}

\title{Constructing Regular Lovelock Black Holes with degenerate vacuum and $\Lambda < 0$, using the gravitational tension. Shadow analysis }
\author{Rodrigo Aros}
\email{raros@unab.cl}
\address{Departamento de Ciencias Fisicas, Universidad Andres Bello, Av. Republica 252, Santiago,Chile} 
\author{Milko Estrada }
\email{milko.estrada@gmail.com}
\affiliation{ Departamento de Física, Facultad de Ciencias, Universidad de Tarapacá, Casilla 7-D, Arica, Chile }
\author{Bastian Astudillo }
\email{b.astudilloespndola@uandresbello.edu}
\address{Departamento de Ciencias Fisicas, Universidad Andres Bello, Av. Republica 252, Santiago,Chile}
\author{R. Prado-Fuentes }
\email{rprado2@santotomas.cl}
\address{Departamento de Ciencias Básica, Facultad de Ciencia, Universidad Santo Tomas, Arica, Chile}

\date{\today}

\begin{abstract}
In \cite{Estrada:2024uuu}, a link between gravitational tension (GT) and energy density via the Kretschmann scalar (KS) was proposed to construct regular black holes (RBHs) in Pure Lovelock (PL) gravity. However, including a negative cosmological constant in PL gravity leads to a curvature singularity \cite{Cai:2006pq}. Here, we choose the coupling constants such that the Lovelock equations admit an $n$-fold degenerate AdS vacuum (LnFDGS), allowing us to construct an RBH with $\Lambda < 0$, where the energy density is analogous to the previously mentioned model. To achieve this, we propose alternative definitions for both the KS and GT. We find that, for mass parameter values greater than the extremal value $M_{\text{min}}$, our RBH solution becomes indistinguishable from the AdS vacuum black hole from inside the event horizon out to infinity. At small scales, quantum effects modify the geometry and thermodynamics, removing the singularity. Furthermore, due to the lack of analytical relationships between the event horizon, photon sphere, and shadow in LnFDGS, we propose a numerical method to represent these quantities.
\end{abstract}

\maketitle

\section{Introduction}

The detection of gravitational waves \cite{LIGOScientific:2016aoc,LIGOScientific:2017ycc} has significantly strengthened the position of General Relativity (GR), even surpassing its already well-established success at the Solar System scale 
\cite{Damour:2007gr}. Nevertheless, some inconsistencies of GR at quantum scales, along with the still unknown nature of dark energy and dark matter, have driven the scientific community to investigate modified theories of gravity. As a first example of these alternative frameworks, the Horndeski scalar-tensor theory \cite{Horndeski:1974wa} generalizes GR by introducing one or more scalar degrees of freedom. This theory is carefully constructed so that its equations of motion prevent the emergence of an unwanted extra degree of freedom, commonly referred to as a 'ghost',which would carry negative energy. As a second example, we consider the $f(R)$ theories of gravity \cite{Nojiri:2006ri}. As noted in reference \cite{Nojiri:2009kx}, this framework offers a gravitational alternative for a unified description of inflation, dark energy, and dark matter, without the need to manually introduce an inflaton field or additional dark components. It is also worth mentioning that black hole solutions have been studied within this theory \cite{delaCruz-Dombriz:2009pzc}.

In line with the previous paragraph, in recent decades, modified gravity theories predicting the existence of extra dimensions have also attracted significant attention. Although several experiments have attempted to test this hypothesis, it has not yet been observed. As a result, any theory that incorporates extra dimensions must be consistent with General Relativity in four dimensions. One such theory is Lovelock gravity. The Lagrangian of Lovelock gravity incorporates higher curvature terms as corrections to the Einstein-Hilbert action \cite{Lovelock:1971yv}. Additionally, Lovelock’s theories respect the core principles of General Relativity; for instance, their equations of motion are of second order. It is worth noting that one particular case of Lovelock theories, namely, Einstein-Gauss-Bonnet gravity, has gained attention in recent years in the context of inflationary models and has been confronted with the GW170817 observational results \cite{Oikonomou:2021kql,Odintsov:2020zkl}.

On the other hand, it is well known that vacuum solutions describing black holes possess a physical singularity, referring to a location where curvature invariants diverge and the tidal forces near a black hole’s singularity become infinite, which can lead to the unbounded stretching of an object, a phenomenon known as spaghettification. One way to address the existence of physical singularities is to define appropriate forms for the matter sources in the energy--momentum tensor such that the central singularity is suppressed. Black holes that are free of a physical singularity are usually referred to as regular black holes (RBHs).

In recent years, the relationship between the gravitational tension associated with the Kretschmann scalar in the vacuum case and the energy density has been investigated to eliminate the existence of singularities. It is well known that an invariant associated with the measurement of tidal forces is the Kretschmann scalar \cite{Bena:2020iyw}. In a four-dimensional, vacuum, spherically symmetric space, this scalar is proportional to $ K \sim {M}^2/r^6 $, leading to infinite tidal forces at the origin. Gravitational field tension in the spherically symmetric case is characterized by the curvature term given by the square root of the Kretschmann scalar of the vacuum solution, $ F \sim \sqrt{K_{\text{Schw}}} \sim \frac{{M}}{r^3} $ in $4D$. This correlation is logical, as the spacetime tension should increase with the mass of the vacuum source \cite{Alencar:2023wyf}.  In this conection, as a way to address the problem of singularities while including an appropriate structure for the energy-momentum tensor, a higher-dimensional model was recently constructed in reference \cite{Estrada:2024uuu} for Pure Lovelock (PL) gravity, where the energy density encodes the gravitational information of the vacuum solution through the Kretschmann scalar. Near the radial origin, where tidal forces and the gravitational tension in the vacuum solution diverge, the tidal forces in the model become finite. Specifically, the energy density is written as:
\begin{equation} \label{probability} 
\rho\sim \exp \left ( - \frac{F_c}{F}\right)       
\end{equation}                             
being $F_c,a$ constants.  

In summary, the model proposed in \cite{Estrada:2024uuu} is such that its energy density encodes the gravitational information of the vacuum solution. Thus, near the central singularity, where the gravitational tension and tidal forces of the vacuum black hole become infinite, the energy density of the model remains finite, suppressing the singularity and giving rise to a regular black hole (RBH). This model coincides with the RBH of Dymnikova \cite{Dymnikova:1992ux} in the four-dimensional case. On the other hand, in a speculative sense, in four dimensions, this way of relating gravitational tension to energy density is analogous to the particle production ratio in the quantum Schwinger effect \cite{DymnikovaS1996}. We can find studies related to this latter topic in the references \cite{Ansoldi:2008jw,Alencar:2023wyf,Estrada:2023pny} . However, this is a topic that requires deeper investigation. 

Since the origins of general relativity, there has been an interest in understanding the role of the cosmological constant. The observed accelerated expansion of the universe can be attributed to a small, yet strictly positive, four dimensional cosmological constant of approximately $\Lambda \sim 10^{-121} \mbox{\,Planck units}\sim 10^{-52} [m^{-2}] \sim 10^{-3} [(\textrm{eV})^4]$ \cite{Barrow:2011zp}. Unfortunately, a positive cosmological constant introduces several technical complications. First, the presence of a positive cosmological constant leads to the existence of a cosmological horizon; thus, it prevents the definition of a proper asymptotic regime and complicates an unambiguous identification of a black hole’s ADM mass. Second, a positive cosmological constant is commonly linked to negative pressure, which implies thermodynamic instability. 

During the last three decades, primarily due to the AdS/CFT conjecture, considering a negative cosmological constant has been of paramount interest in theoretical physics. From a theoretical point of view, considering asymptotically AdS spaces has advantages, such as the construction of a proper action principle —one that admits the definition of an asymptotic boundary where well-defined boundary conditions can be imposed and where unambiguous conserved charges can be computed.

Despite the above mentioned issues regarding physical problems addressed by including a negative cosmological constant, there has been little speculation in the literature about a possible numerical negative value for this constant. For example, references \cite{Zakharov:2014pca,Zakharov:2018awx}, using observations of stars near the Galactic Center, estimate a negative value for the cosmological constant: in \cite{Zakharov:2014pca}, for anti–de Sitter spacetimes corresponding to a negative $\Lambda$ term of approximately $-0.4 \times 10^{-20} \text{ m}^{-2}$. The authors claim that this value is consistent with some ideas within the framework of a certain class of multidimensional string models. In \cite{Zakharov:2018awx}, also using observations of stars near the Galactic Center, the orbital precession in the Reissner–Nordström–AdS solution is studied, obtaining values of $-4.68 \times 10^{-39} \text{ cm}^{-2} < \Lambda < 0$. The authors state that their constraints on the cosmological constant are much weaker not only than their cosmological estimates but also than the constraints obtained from Solar System data.

In the context of this work, that is, within Lovelock theories, some authors assume that the coupling constants of the theory could be of the order of Planck scale quantities \cite{Chernicoff:2016uvq}. This makes sense if one considers that the theory might only be testable at high energies. As we will see in detail below, the zeroth-order coupling constant is proportional to the cosmological constant. Under this idea, a negative cosmological constant would be of the order of powers of the Planck length, $\Lambda \sim -\ell_p^{-2n}$. That is, an AdS radius on the order of the Planck length. However, there are also works that define intermediate energies as those which are weak enough to be testable at energy scales below the Planck scale, but still high enough to feel the effects of the higher-curvature terms in Lovelock theory \cite{Camanho:2014apa}. Under this latter assumption, $\Lambda \sim -\bar{\ell}^{-2n}$, with $\bar{\ell} > \ell_p$. That is, an AdS radius sufficiently larger than the Planck length, in order to satisfy what has been mentioned above.

When higher dimensions ($d>4$) are considered, several families of gravitational theories emerged that satisfy the same conditions of General Relativity, such as second-order differential equations. Arguably, the simplest of these families is the $d$-dimensional Lovelock gravity, which corresponds to the addition of all of the lower-dimensional topological densities. General relativity corresponds to the simplest case.    

A special case of Lovelock theories is Pure Lovelock theory (PL) \cite{Cai:2006pq,Dadhich:2012cv}. The latter theory mentioned considers {\it only a single} term in the Lagrangian. Related to those discussed above, a disadvantage of PL theory is that considering the presence of a negative cosmological constant in a spherically symmetric spacetime leads to the appearance of a curvature singularity. This singularity is such that, although the metric tensor is regular for a radius greater than the event horizon, the Ricci and Kretschmann invariants diverge at a point within this location. This issue has not been extensively investigated; therefore, it may be related to a potential breakdown of spacetime. Thus, studying the recent analogy between gravitational tension and energy density, as carried out in \cite{Estrada:2024uuu}, appears to be unfeasible within the framework of PL theory in the presence of a negative cosmological constant for the construction of RBH solutions.

To address the issue above—namely, the apparent impossibility of studying the recent analogy between gravitational tension and energy density in pure Lovelock theory under the influence of a negative cosmological constant—this work will take the following into account: One potential drawback of generic Lovelock gravity is the existence of multiple ground states, meaning more than one solution with constant curvature spaces, or equivalently, more than one potential effective cosmological constant \cite{Camanho:2011rj}. These effective cosmological constants can be complex numbers, which can lead to instability of the ground states under dynamical evolution. To address this issue, one can choose the coupling constant in such a way that the equations of motion roughly take the form 
$(R-\Lambda)^n=0$ \cite{Crisostomo:2000bb}. In this case, there is a single ground state, but it is n-fold degenerate with constant curvature. This scenario is referred to as Lovelock with an n-fold degenerate ground state (LnFDGS). This is also known as the degenerate vacuum. In the context of this work, this coupling constant structure allows for the observation of a single negative effective cosmological constant, which, among other things, has facilitated the development of techniques for determining conserved charges \cite{Aros:2019quj}. This also seems to make sense in the framework of the AdS/CFT correspondence, as two or more negative cosmological constants would lead to different asymptotic boundary structures for the same background.  See some  recent applications of LnFDGS theory in references \cite{Xu:2019krv,Aros:2019quj,Estrada:2019qsu,Aros:2019vpu,Estrada:2021kuj,Aros:2023tbh}

PL has significant differences compared to the LnFDGS theory \cite{Aros:2019quj,Crisostomo:2000bb}. While PL considers only a single term from the Lovelock series, LnFDGS includes all the terms from this series up to $n = [d/2]$, where the square brackets represent the floor function. Moreover, PL cannot be factored into several equivalent negative effective cosmological constants, such as LnFDGS. Remarkably, LnFDGS does not exhibit a curvature singularity of the type mentioned earlier. The latter offers the advantage of displaying a well-defined asymptotically AdS structure. In the context of this work, the absence of the aforementioned pathology in LnFDGS enables the construction of RBH solutions with a negative cosmological constant, allowing for the testing of the physical consequences of its presence from both geometric and thermodynamic perspectives.

On the other hand, it is well known that light is not directly observable at the event horizon of a black hole. Instead, what can be detected is the so-called shadow of a black hole, a dark region surrounded by light, resulting from gravitational lensing around the event horizon. One of the major achievements of the Event Horizon Telescope (EHT) collaboration has been the imaging of the shadows of the supermassive black holes M87 \cite{EventHorizonTelescope:2021srq} and Sgr A \cite{EventHorizonTelescope:2022xqj}. It is worth mentioning that, given the theoretical interest driven by the presence of a negative cosmological constant, the study of shadows in an asymptotically AdS regime has also been explored in some references (see, for example, \cite{Belhaj:2021rae,Hamil:2023zeb}). In this context, there has also been strong motivation, from a theoretical perspective, to study shadows in extra-dimensional scenarios. See, for example, reference \cite{Nozari:2024jiz} for Scalar-Tensor-Vector theories, and reference \cite{Singh:2023ops} (and references therein) for rotating black hole shadows.  See also \cite{Vagnozzi:2019apd}. As previously mentioned, the study of the physical properties of black holes in Lovelock gravity has attracted considerable attention in recent years \cite{Aros:2019quj,Xu:2019krv,Estrada:2019qsu,Aros:2019vpu,Estrada:2021kuj,Aros:2000ij,Bueno:2024dgm}. Along this line, it is natural to investigate the physical properties of their shadows. It is worth noting that in reference \cite{Paithankar:2023ofw}, analytical relations were established for the photon sphere radius and the shadow radius of singular vacuum black holes in pure Lovelock (PL) gravity. However, for most specific cases within Lovelock theory, such analytical results are not attainable. Therefore, it is of physical interest to develop a numerical methodology for representing black hole shadows.

Regarding the points above, in this work, we will provide a methodology to study the analogy between gravitational tension and energy density, aiming to construct regular black hole solutions in LnFDGS gravity in the presence of a negative cosmological constant. It is worth mentioning that directly relating the gravitational tension to the Kretschmann scalar of the vacuum LnFDGS solution, in the same way as for PL theory, is complicated because several additional terms would appear in the gravitational tension. Due to this, unlike \cite{Estrada:2024uuu}, we will define an alternative version of the Kretschmann scalar suitable for the LnFDGS AdS theory, and consequently, a redefinition of the gravitational tension.  We will test how the presence of the negative cosmological constant influences the obtained RBH solution, both in terms of its geometric and thermodynamic properties. Additionally, since it is not possible to find analytical relationships between the event horizon, the photon sphere radius, and the shadow size in LnFDGS, we propose a method to numerically and graphically obtain these relationships and analyze their physical behavior. 

\section{A brief revision of Lovelock gravity with \textit{n}-fold degenerated ground state}

Given that this family of theories is well known, it is worthwhile to recall only the relevant features for the discussion. First, the $d$-dimensional Lovelock (\textbf{LL}) Lagrangian  is given by:

\begin{equation}\label{LL}
  L = \sum_{p=0}^{n=[d/2]} \frac{1}{2^p} \alpha_p \delta^{\nu_1\ldots \nu_2p}_{\mu_1\ldots \mu_2p} R^{\mu_1\mu_2}_{\hspace{2ex}\nu_1\nu_2}\ldots R^{\mu_{2p-1}\mu_{2p}}_{\hspace{2ex}\nu_{2p-1}\nu_{2p}}
\end{equation}
where $R^{\mu \alpha}_{\hspace{2ex}\nu \beta}$ is the Riemann tensor, $[X]$ stands for the integer part of $X$ and $\{\alpha_p\}$ is a set of dimensionful constants. As mentioned in the introduction, since they are dimensionful coupling constants, they can be defined such that they have units of $\alpha_p \sim [\text{length}]^{2p - d}$ \cite{Garraffo:2008hu}. Thus, it is possible to define $ \alpha_0 \sim \Lambda  \alpha_n$, with $\Lambda$ having units of $\Lambda \sim [\text{length}]^{-2n}$. Therefore, for $n = 1$, corresponding to the Einstein-Hilbert case, we have $\Lambda \sim [\text{length}]^{-2}$. The \textbf{LL} equations of motion can be written as
\begin{equation}\label{LLEOM}
   \mathcal{G}^{\alpha}_{\hspace{1ex}\beta}= \sum_{p=0}^{n=[d/2]} \frac{d-2p}{2^p} \alpha_p \delta^{\alpha\nu_1\ldots \nu_2p}_{\beta\mu_1\ldots \mu_2p} R^{\mu_1\mu_2}_{\hspace{2ex}\nu_1\nu_2}\ldots R^{\mu_{2p-1}\mu_{2p}}_{\hspace{2ex}\nu_{2p-1}\nu_{2p}}.
\end{equation}

Here $\mathcal{G}^{(n)}_{\alpha \beta}$ denotes an $n$-th order generalization of the Einstein tensor, which is influenced by the order on the Riemann tensor of the Lagrangian $L$. For instance, $\mathcal{G}^{(1)}_{\alpha \beta}$ represents the Einstein tensor associated with the Ricci scalar (with the Einstein-Hilbert theory being a specific case of Lovelock theory), the $\mathcal{G}^{(2)}_{\alpha \beta}$ corresponds to the Lanczos tensor related to the Gauss-Bonnet Lagrangian and $\mathcal{G}^{(3)}_{\alpha \beta}$ corresponds to the cubic tensor.

It's straightforward to notice that not even in a vacuum can these equations of motion be solved for an arbitrary set of $\alpha$'s. Not even the ground-states, \textit{i.e.} the locally constant curvature solutions, can be obtained in $d>8$ for a general set of $\alpha$'s. The problem of obtaining those constant curvature solutions reduces to getting the roots of a polynomial of order $[d/2]$, which, as it is well known, is not unattainable for $[d/2]>4$. Moreover, those roots can be complex numbers with nonvanishing imaginary parts. A way to avoid that discussion is to fix the asymptotic behavior to a single asymptotically locally flat or AdS behavior. 

As mentioned in the introduction, Lovelock theory can be factorized in several effective cosmological constants \cite{Camanho:2011rj}. In this connection, for $\alpha_p=0$ from $p > I$, the vacuum equations of motion can be written as \cite{Aros:2019quj}:

\begin{equation}\label{LovelockFactorizado}
 \mathcal{G}^{\mu}_{ (LL)\hspace{1ex} \nu} \propto \delta_{\mu_1 \nu_1 \ldots \mu_{I}\nu_I \nu}^{\alpha_1\beta_1 \ldots \alpha_{I}\beta_I\mu} (R^{\nu_1 \mu_1}_{\hspace{2ex}\alpha_1 \beta_1} + \kappa_1 \delta^{\mu_1 \nu_1}_{\hspace{2ex}\alpha_1 \beta_1})\ldots (R^{\nu_I \mu_I}_{\hspace{2ex}\alpha_I \beta_I} + \kappa_I \delta^{\mu_I \nu_I}_{\hspace{2ex}\alpha_I \alpha_I}) =0. 
\end{equation}

This shows, as expected, that the Lovelock gravity can be factorized in several ground states of constant curvature.
In this work, we aim to complete the analysis considering asymptotically locally AdS spaces, \textit{i.e.}, spaces satisfying
\begin{equation}\label{ALAdSCondition}
 \left.R^{\mu \alpha}_{\hspace{2ex}\nu \beta}\right|_{x \rightarrow {\text{ALAdS region}}} \longrightarrow -\frac{1}{l^2} \delta^{\mu \alpha}_{\nu \beta}
\end{equation}
In the simplest cases where Eq. (\ref{ALAdSCondition}) is satisfied, the vacuum is $n$-fold degenerate, and the EOMs in Eqs. (\ref{LLEOM}) and \eqref{LovelockFactorizado}, after an appropriate choice of the coupling constants \cite{Crisostomo:2000bb,Aros:2000ij}, take the simple form
\begin{equation}\label{LLEOMAdS}
   \mathcal{G}^{\alpha}_{\hspace{1ex}\beta}= \alpha_0 \delta^{\alpha\nu_1\ldots \nu_{2n}}_{\beta\mu_1\ldots \mu_{2n}} \left(R^{\mu_1\mu_2}_{\hspace{2ex}\nu_1\nu_2} + \frac{1}{l^2}\delta^{\mu_1\mu_2}_{\nu_1\nu_2}\right)\ldots \left(R^{\mu_{2p-1}\mu_{2p}}_{\hspace{2ex}\nu_{2n-1}\nu_{2n}}+ \frac{1}{l^2}\delta^{\mu_{2n-1}\mu_{2n}}_{\nu_{2n-1}\nu_{2n}}\right)
\end{equation}
with $0<n <[d/2]$. 

The static vacuum solution of (\ref{LLEOMAdS}) was obtained in \cite{Crisostomo:2000bb,Aros:2000ij}. In Schwarzschild coordinates, this is given by
\begin{equation}\label{SchwCoordinates}
  ds^2 = -f(r) dt^2 + \frac{1}{f(r)} dr^2 + r^2 d\Omega_{d-2}
\end{equation}
where $d\Omega_{d-2}$ corresponds to the transversal section of a $(d - 2)$-sphere. The vacuum solution is
\begin{equation}\label{ACTZSol}
  f(r) = 1 + \frac{r^2}{l^2} - \left(\frac{2M}{r^{d-2n-1}}\right)^{\frac{1}{n}}.
\end{equation}
It must be stressed that $(X)^{1/n}$ stands for any of the $n$ different roots of (X). It is straightforward to check the presence of a singularity at $r=0$. Nonetheless, this is not enough, and in general, what characterizes the interesting physical solutions is that they satisfy $f(r)^2 \in \mathbb{R}$, for $r>0$. If $\exists \hspace{1ex} r^*$ such $f(r^*)^2=0$ this defines a Killing horizon and thus the line element above Eqs.(\ref{SchwCoordinates}) and \eqref{ACTZSol} describe a black hole geometry. One can notice how similar is this solution to the corresponding Schwarzschild solution \footnote{Nonetheless, at least there is significant difference.  For $n$ even integer, since there are always two $\pm$ real roots in the expression, $M<0$ can be considered allowed.}. 

It is worth stressing that, as mentioned in the introduction, the structure of the equations of motion leading to the last solution is quite different from that of PL, as PL considers only a single term in the Lovelock series. At the same time, LnFDGS includes all the terms of the mentioned series up to $n = [d/2]$ under the particular choice of coupling constants previously mentioned. This choice allows the equations of motion to be factorized into AdS vacua, as described above (which is not possible for PL).

\section{About the nature of the matter source proposed for Lovelock with \textit{N}-fold degenerated ground state AdS.}

Given the aforementioned differences between PL and LnFDGS theories, providing a form for the energy density analogous to the one described earlier is not straightforward. This is because several additional terms would appear when directly substituting the Kretschmann scalar of the vacuum LnFDGS solution. To address this, we will provide some of the necessary ingredients. First, we will define gravitational tension through the gauge curvature tensor $F_{\mu\nu}$, which is the Poincaré (group) curvature. Second, we will define an appropriate tensor for LnFDGS, which we will refer to as the Poincaré AdS-like curvature. From this, we will provide an alternative version of the Kretschmann scalar associated with gravitational tension. Finally, we will describe, from a physical perspective, the resulting structure of the energy density.

\subsection{Defining gravitational tension through the gauge curvature and a glimpse of the analogy with the Schwinger effect.}
Although the analogy between gravitational tension and the production ratio in the Schwinger effect requires further study, this work provides some insights into the subject, allowing us to speculate. As discussed in \cite{Ansoldi:2008jw,Dymnikova:1992ux}, one could argue that the Schwinger effect should play a role in one of two ways. Firstly, it would give rise to corrections to the vacuum expectation value of the fields. These corrections can be expressed in terms of powers of the Riemann tensor \cite{Birrell:1982ix}. Furthermore, gravitation itself should self-correct due to the presence of singularities. It is expected that these effects cannot be separated for large curvature. To address any of these approximations, after considering Eq. (\ref{probability}), one can recall that in asymptotically (Riemannian) flat spaces (i.e., with a vanishing cosmological constant), one analogy to the gauge curvature $F_{\mu\nu}$ is the Poincaré (group) curvature. In this way, we can also interpret the gravitational tension as $F = \sqrt{F_{\mu\nu}F^{\mu\nu}}$, where

\begin{equation} F^{\text{P}}_{\mu\nu}=e^{a}_{\hspace{1ex}\lambda}e^{b}_{\hspace{1ex}\rho} R^{\lambda\rho}_{\hspace{2ex} \mu\nu} J_{ab} + e^{a}_{\hspace{1ex}\rho} T^{\rho}_{\hspace{1ex}\mu\nu} P_{a}
\end{equation}
where $(J_{ab},P_c)$ span the Poincare algebra. Here $ R^{\lambda\rho}_{\hspace{2ex} \mu\nu}$ is the curvature tensor and $T^{\rho}_{\hspace{1ex}\mu\nu}$ the torsion tensor. However, since on a Riemannian manifold the torsion tensor vanishes and the curvature tensor is the Riemann tensor, the Schwinger effect could to be characterized by the Kretschmann scalar \cite{Ansoldi:2008jw,DymnikovaS1996,Bena:2020iyw}
\begin{equation}\label{Kretschmann}
F_{\mu\nu}F^{\mu\nu} \triangleq K =R^{\lambda\rho}_{\hspace{2ex} \mu\nu}R^{\mu\nu}_{\hspace{2ex} \lambda\rho}.
\end{equation}

Now, as mentioned earlier, in order to propose an energy density, one can observe that for the four-dimensional Schwarzschild solution, the Kretschmann scalar is proportional to $ K \sim 1/r^6 $, and thus, by substituting it into Eq. \eqref{probability}, the energy density proposed by Dymnikova is reproduced.

Before proceeding, it is worth mentioning that the prescription above has been further explored in many works, see \cite{Ansoldi:2008jw,DymnikovaS1996,Estrada:2023pny,Alencar:2023wyf}. Essentially, the results are regular black hole and wormhole solutions. In \cite{Wondrak:2023zdi,Chernodub:2023pwf}, the mentioned relationship between the Kretschmann scalar and the Schwinger effect was explored in a different (gravitational) context.

\subsection{An alternative definition of the Kretschmann scalar for our anti-de Sitter space.}
To extend the framework to asymptotically (locally) Anti de Sitter spaces, it is enough to consider a $SO(d-1,2$) curvature \cite{Hassaine:2016amq} instead of the Poincare one. One option is \cite{Hassaine:2004pp}
\begin{equation}\label{AdSCurvature}
  F^{\text{AdS}}_{\mu\nu} = e^{a}_{\hspace{1ex}\lambda}e^{b}_{\hspace{1ex}\rho}\left(R^{\lambda\rho}_{\hspace{2ex} \mu\nu} + \frac{1}{l^2} \delta^{\lambda\rho}_{\mu\nu}\right)J_{ab} + e^{a}_{\hspace{1ex}\rho} T^{\rho}_{\hspace{1ex}\mu\nu} J_{a},
\end{equation}
In this work, for simplicity, we will refer to the tensor $F^{\text{AdS}}_{\mu\nu}$ as Poincaré AdS-like curvature,
where $l$ is called the AdS radius and $J_{ab},J_{a}$ are a set of generators of $SO(d-1,2)$. Since the discussion is for Riemannian manifold, where $T^{\rho}_{\hspace{1ex}\mu\nu}=0$, in order to simplify the calculations, one can define an adequate Kretschmann scalar $K'$
\begin{equation} \label{Kprima}
(F_{\mu\nu})^{\text{AdS}}(F^{\mu\nu})^{\text{AdS}} \triangleq K'= \left(R^{\lambda\rho}_{\hspace{2ex} \mu\nu} + \frac{1}{l^2} \delta^{\lambda\rho}_{\mu\nu}\right)
\left(R_{\lambda\rho}^{\hspace{2ex} \mu\nu} + \frac{1}{l^2} \delta_{\lambda\rho}^{\mu\nu}\right) = K + \frac{8}{l^2} R + \frac{2}{l^4}(d-1).
\end{equation}
As done above, one can evaluate $K'$ on the four-dimensional Schwarzschild-AdS solution, which yields
\begin{equation}
  K' \sim \frac{1}{r^6}.
\end{equation}
This leads to the same energy density, see Eq.(\ref{probability}), of the $\Lambda=0$ case. This must be expected as for $r\approx 0$ the effects of the different asymptotical structure should be irrelevant.

\subsection{The proposed model}

As mentioned earlier, to define the energy density in \eqref{probability}, we will associate the gravitational tension with the value of our definition provided above for the Kretschmann scalar for the vacuum AdS LnFDGS solution. From equations \eqref{LLEOMAdS} and \eqref{Kprima}
\begin{equation}
  K'=  \left(\frac{d^2}{dr^2} g(r)\right)^2 + \frac{2(d - 2)}{r^2}\left(\frac{d}{dr}g(r)\right)^2 + \frac{2(d - 2)(d - 3)}{r^4}\left(g(r)\right)^2
\end{equation}
where
\begin{equation}
g(r) = 1 + \frac{r^2}{l^2}- f(r).
\end{equation}
Now, evaluating, using \eqref{ACTZSol}, yields
\begin{equation}
    K' =\left(\frac{(d - n - 1)^2}{n^4} + 2(d - 2)\left(\frac{1}{n^2} + (d - 3)\right)\right)  \frac{(2M)^{\frac{2}{n}}}{r^{\frac{2}{n}(d-1)}}
\end{equation}
Building on the idea previously described, where gravitational tension is proportional to the square root of the Kretschmann scalar of the vacuum solution
\begin{equation}
    F \sim \sqrt{K'} \sim \frac{{M}^{\frac{1}{n}}}{r^{\frac{d-1}{n}}}
\end{equation}
In this way, the steps previously taken, namely: defining a tensor analogous to the gauge curvature, which we have called as Poincaré AdS-like curvature and providing an alternative definition for the Kretschmann scalar, now allow us to model an energy density. Thus, in analogy with \eqref{probability}, it can be defined the energy density
\begin{equation} \label{ModeloDensidad}
\rho = A \exp \left ( - \frac{r^{(d-1)/n}}{a^{(d-1)/n}} \right )
\end{equation}
where, for simplicity, the constant $A$ has been adjusted to:
\begin{equation} \label{ConstanteAdensidad}
    A= \frac{d-2}{n} \frac{M}{a^{d-1}/(d-1)}
\end{equation}
where $a$ is an arbitrary constant. It is worth mentioning that this density model satisfies what was described earlier. While the gravitational tension associated with the tidal forces of the vacuum LnFDGS AdS solution diverges near the origin, the energy density encodes the information of the aforementioned case in such a way that, at the origin, the density takes a finite value, which, as we will see below, is associated with the suppression of the usual singularity present at the radial origin.

\section{A new static regular solution in Lovelock with $n$ fold degenerated ground state AdS }

The energy-momentum tensor corresponds to a neutral perfect fluid:
\begin{equation}
T^\mu_\nu=\mbox{diag}(-\rho,p_r,p_\theta,p_\theta,...),
\end{equation}
On the one hand, it is well known that this form of the metric imposes the condition \(\rho = -p_r\). To obtain the regular solution it is enough to solve (\ref{LLEOMAdS})
\begin{equation}\label{LLEOMAdSReg}
   \mathcal{G}^{\alpha}_{\hspace{1ex}\beta}=  T^{\alpha}_{\hspace{1ex}\beta}
\end{equation}
The components $(t,t)$ and $(r,r)$ of the equations of motion take the form:
\begin{equation} \label{ttcomponente}
 \frac{d}{dr} \left (r^{d-1} \left( \bar{f}(r) + \frac{1}{l^2} \right)^{n}   \right ) = \frac{2}{d-2} r^{d-2} \rho ,
\end{equation}
where $l$ corresponds to the AdS radius, which is related to the cosmological constant such that $\Lambda = -\frac{(d-1)(d-2)}{2l^{2}}$ and where
\begin{equation}
    \bar{f}(r) =  \frac{1 -f(r)}{r^2}
\end{equation}
On the other hand, due to the transverse symmetry, we have $p_\theta = p_\phi = p_t = \ldots$ for all the $(d-2)$ angular coordinates. Thus, using the aforementioned condition $\rho = -p_r$, the conservation law $T^{\alpha \beta}_{;\beta} = 0$ gives:
\begin{equation} \label{conservacion1}
p_t = -\frac{r}{d-2} \rho' - \rho
\end{equation}

Using the static ansatz (\ref{SchwCoordinates}) with
\begin{equation} \label{SolucionRegular}
f(r) = 1 + \frac{r^2}{l^2} - \left(\frac{2m(r)}{r^{d-2n-1}}\right)^{\frac{1}{n}}
\end{equation}
the direct integration yields
\begin{equation} \label{SolucionFuncionMasa}
    m(r)= \frac{1}{d-2}  \int \rho r^{d-2} dr
\end{equation}
We replace the energy density in the last equation, using equations \eqref{ModeloDensidad} and \eqref{ConstanteAdensidad}, obtaining the mass function

By choosing $M \cdot(n-1)!$ as integration constant, the mass function is given by

\begin{eqnarray} \label{FuncionMasaRegular}
m(r) &=& M \left ( (n-1)! - \Gamma \left [ n \, , \, \frac{r^{(d-1)/n}}{a^{(d-1)/n}} \right ]  \right )
\end{eqnarray}
In the expression above, we have chosen $M \cdot (n - 1)!$ as the integration constant. As we will see below, this choice allows the function to behave as a de Sitter core close to the origin, and thus, the solution is regular.

It is worth noting that the coordinate $r$ defines the regions of interest. While $r \rightarrow \infty$ corresponds to the asymptotic region, $r \rightarrow 0$ must define the center of the solution. Below, we will analyze the regions associated with the event horizons present in the solution.

\subsection{Structure of horizons}

We will begin by mentioning that the mass parameter, at which our solution, equations \eqref{SolucionRegular} and \eqref{SolucionFuncionMasa} vanishes, is given by the following expression:

\begin{equation} \label{EqParametroMasaAgosto}
  M(\bar{r})=  \frac{\bar{r}^{-1 + d - 2n} \left( 1 + \frac{\bar{r}^2}{l^2} \right)^n}{2 \left( (-1 + n)! - \Gamma\left[n, \left( \frac{\bar{r}}{a} \right)^{\frac{d - 1}{n}} \right] \right)}
\end{equation}
where $\bar{r}$ is such that $f(r = \bar{r}) = 0$.

\begin{center}
\begin{figure}[h]
  \includegraphics[width=4.5in]{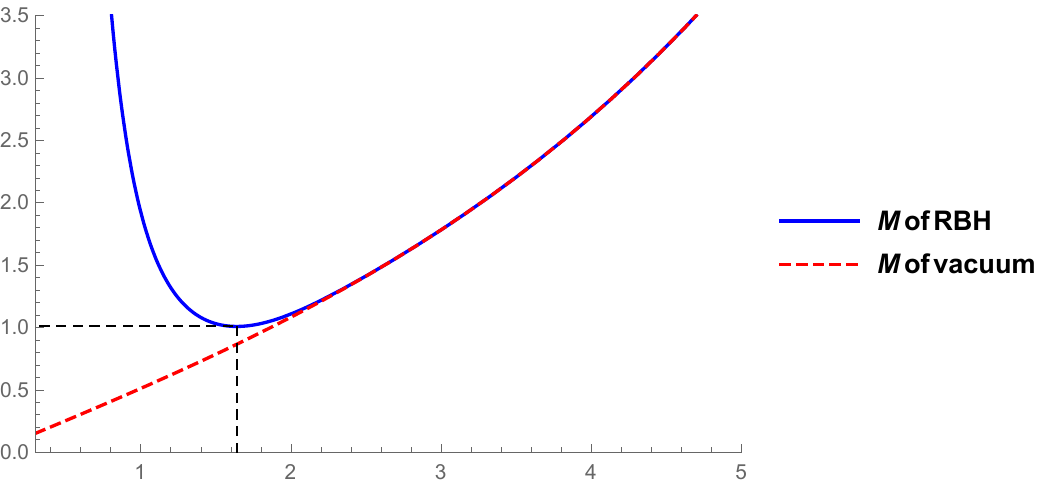}
  \caption{In horizontal axis $\bar{r}$. In vertical axis: $M$ of our RBH is shown in blue and $M$ in the vacuum case \cite{Aros:2000ij} in dashed red. We have used $d = 6,n = 2,l = 10,a = 1$. We can check that this behavior is generic by using other parameter values. The decreasing part of the blue curve corresponds to the inner horizon of the RBH solution, while the increasing part represents the event horizon. The minimum point marks the extremal radius. The red curve represents the vacuum AdS solution. For a value slightly greater than the extremal radius, both solutions become indistinguishable for radial values equal to or greater than the event horizon.} \label{TwoMasses}
\end{figure}
\end{center}

In Figure \ref{TwoMasses}, we can observe the generic behavior of the mass parameter as a function of the radial values where the function $f(r)$ vanishes. The behavior in the vacuum LnFDGS AdS case \cite{Crisostomo:2000bb, Aros:2000ij} is shown in dashed red, while the behavior of our RBH LUV AdS solution with our matter sources is shown in blue. Thus, the regions where $M$ decreases correspond to the values of the inner horizon $\bar{r}=r_-$, and the regions where $M$ increases correspond to the values of the black hole horizon $\bar{r}=r_+$ in our solution. Therefore, for $M > M_{\text{min}}$, there exist an inner horizon $r_-$ and a black hole horizon $r_+$ for the same value of the parameter $M$. The value $M_{\text{min}}$ corresponds to the point where the inner and black hole horizons coincide, i.e., an extremal black hole. 

A remarkable feature is that, under a small rightward deviation in the value of the extremal radius $r_{\text{ext}} = r_- = r_+$, that is, $\bar{r}=r_* > r_{\text{ext}}$ (which corresponds to a small upper-right deviation in the value of $M$ with respect to $M_{\text{min}}$), our regular AdS solution and its vacuum AdS counterpart become indistinguishable, as the dashed red and blue curves also become indistinguishable. More specifically, for a value slightly above $M_{\text{min}}$, both the vacuum solution and the regular solution share the same event horizon radius. In this case, the two solutions can only be distinguished inside the event horizon.

For values on the vertical axis smaller than $M_{\text{min}}$, our solution exhibits a regular geometry without horizons. It is worthwhile to stress that, unlike the vacuum LnFDGS AdS solutions in \cite{Crisostomo:2000bb,Aros:2000ij}, due to the absence of singularities, regular solutions do not need to be black holes, i.e., they do not require the presence of a horizon to be well defined. In other words, regular solutions are not constrained by the \textit{cosmic censorship} hypothesis.

If we take the Planck length as a reference on the horizontal axis of the figure, such that the extremal radius is $r_{\text{ext}} \sim 1.6 , \ell_p$ and $M_{\text{min}} \sim \ell_p$, then for values slightly greater than these, $r_* > r_{\text{ext}}$, both solutions become indistinguishable. In this scenario, physics at Planck scales becomes particularly interesting, as the nature of the matter sources contributes to the suppression of the singularity. 

As a consequence of the above, we also note that for values considerably larger than the Planck scales, such as the observed values of mass and radius of rotating black holes observed in LIGO GW150914, about 35 solar masses \cite{LIGOScientific:2016wyt}, both the vacuum and regular solutions are also indistinguishable. Thus, it becomes intriguing to observe the properties of black holes at scales much smaller than those that have been observed to date.

It is worth discussing the potential case where $r_+ \gg a$, in which, as a consequence of the previous discussion, the Schwarzschild LnFDGS AdS solution \cite{Crisostomo:2000bb, Aros:2000ij} becomes indistinguishable from our regular black hole (RBH) solution with LnFDGS AdS. This scenario makes sense if one considers the case in which the parameter $a$ is of the order of the Planck length, $a \sim \ell_p$, and the radius is of the order of the rotating black hole observed in LIGO GW150914 \cite{LIGOScientific:2016wyt}, $r_+ \sim 103$ km. In this way, the parameter $a$ can be interpreted essentially as the length scale at which quantum corrections become not only non-negligible but also relevant. Depending on the model considered, this length scale can be as small as the Planck length.

\subsection{Merging zone}
\begin{center}
\begin{figure}[h]
  \includegraphics[width=4.5in]{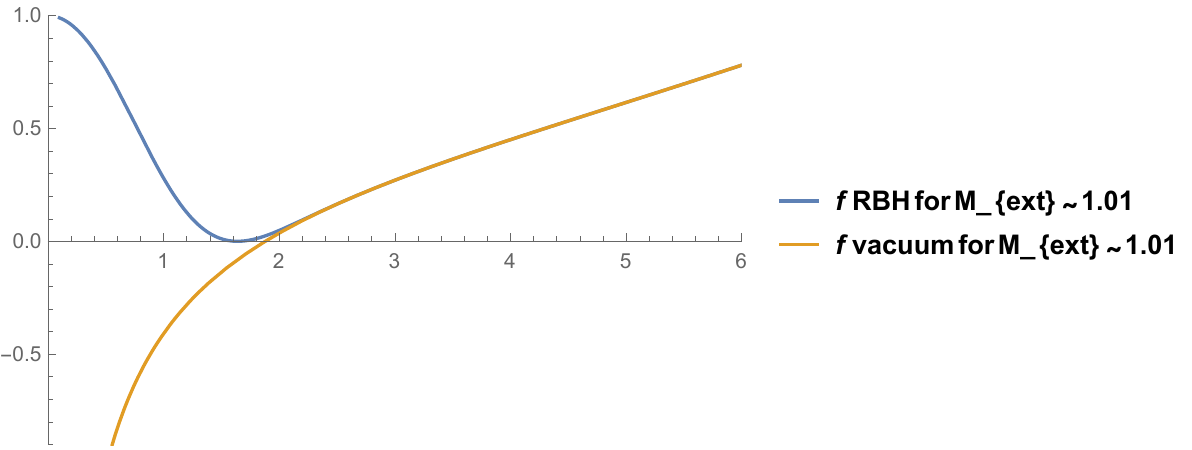}
   \includegraphics[width=4.5in]{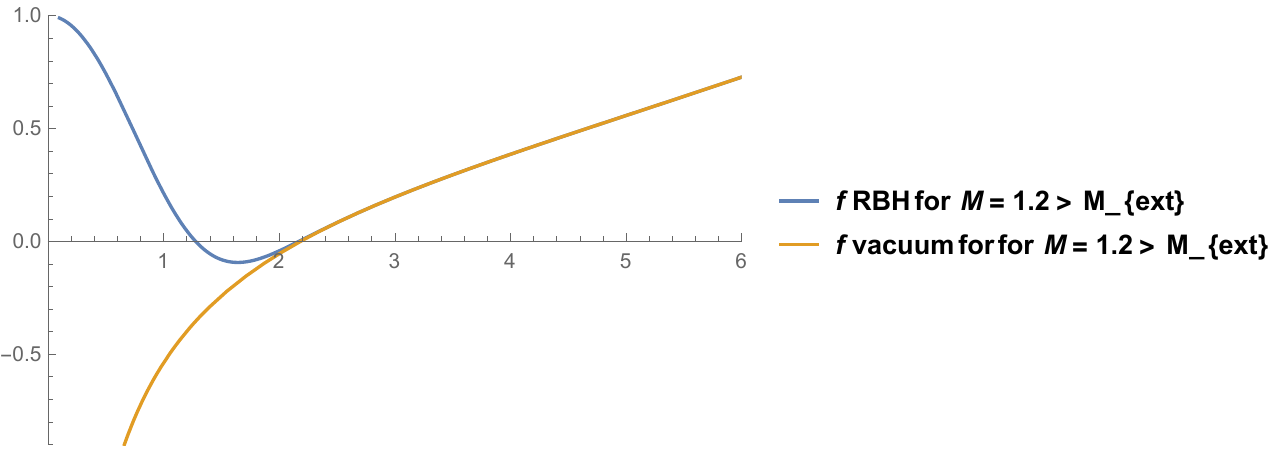}
    \includegraphics[width=4.5in]{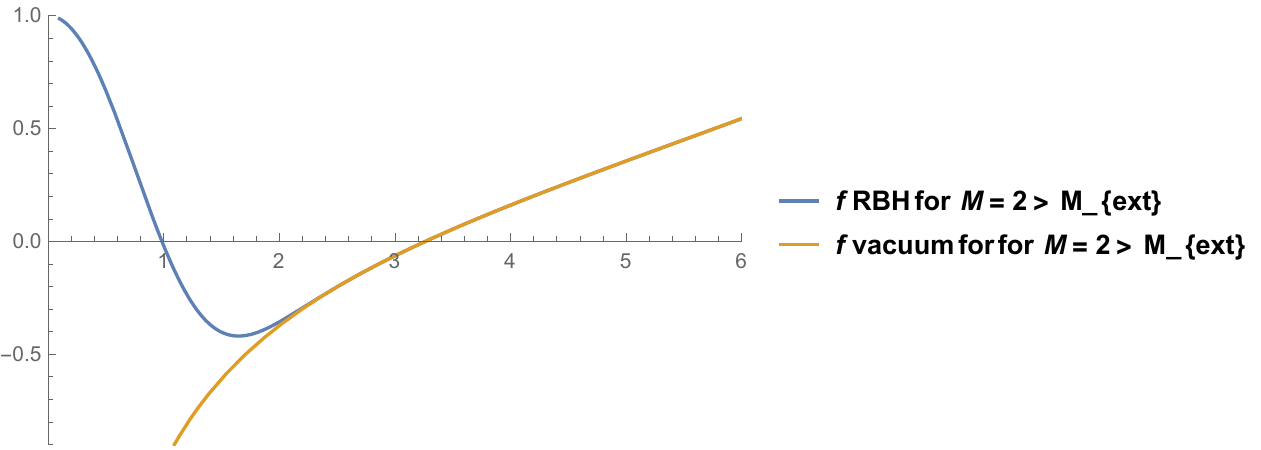}
  \caption{On the horizontal axis, $r$. On the vertical axis, the function $f(r)$ for $M = M_{\text{ext}} \approx 1.01$, $M = 1.2 > M_{\text{ext}}$, and $M = 2 > M_{\text{ext}}$ in the first, second, and third panels, respectively. We have used $d = 6$, $n = 2$, $l = 10$, $a = 1$. We can check that this behavior is generic for other parameter values. We observe that for mass parameter values greater than the extremal mass, $M > M_{\text{ext}}$, both solutions become indistinguishable from within the event horizon.} \label{ZonaMerging}
\end{figure}
\end{center}

 A remarkable feature of our regular solution is the existence, in general, of a radial coordinate value at which the regular solution converges toward the vacuum LnFDGS AdS solution found in Refs.\cite{Crisostomo:2000bb,Aros:2000ij}, which corresponds to Eq.(\ref{ACTZSol}). However, analytically estimating the precise radial value beyond which our regular AdS solution and its vacuum AdS counterpart become indistinguishable is difficult. Numerically, we observe in the first panel of Figure~\ref{ZonaMerging} that, for a value of the mass parameter $M_{min}$, associated with the existence of an extremal horizon $r_{ext}$, both solutions become indistinguishable for a radial value greater than the extremal horizon. In the second panel, we observe that for mass values greater than $M_{min}$, corresponding to an event horizon $r_* = r_+ > r_{ext}$, both solutions become indistinguishable before reaching the mentioned event horizon. It is straightforward to verify that the behavior shown in this figure is generic for other parameter choices. In this way, the obtained solution is such that, for a mass parameter value slightly greater than $M_{\text{min}}$ (associated with an extremal radius), our RBH solution LnFDGS becomes indistinguishable from its vacuum counterpart LnFDGS \cite{Crisostomo:2000bb,Aros:2000ij} inside the event horizon, from a radial value $r = r_* < r_+$ up to infinity. If $r_*$ is of the order of the Planck scale, the geometric differences between both solutions, particularly the suppression of the singularity, would occur at quantum scales.

\subsection{Behavior near $r=0$ region}

One must analyze the region near $r = 0$, which essentially separates the regular solution from its black hole counterpart. At $r = 0$, the absence of any singularity can be demonstrated. This becomes evident simply by taking the limit.
\begin{equation}\label{NonSingular}
  \lim_{r\rightarrow 0} f(r) \rightarrow 1 - \left(\left(\frac{2M}{n}\right)^{\frac{1}{n}} - \frac{a^2}{l^2}\right) \frac{r^2}{a^2}.
\end{equation}
n principle, this defines a (nearly) (Anti) de Sitter or flat region near $r = 0$. However, the presence of the term $a^2 / l^2$ can be observed, and therefore, given that in a realistic physical setup $a / l \ll 1$, where $l$ represents the AdS radius, even for small masses $M$, a de Sitter-like region is expected near $r = 0$.

From the last equation, it is straightforward to check that the Kretschmann invariant takes a finite value near the origin:

\begin{equation}
    \displaystyle \lim_{r \to 0} K = \frac{2(d^2-5d+8)}{a^4} \left ( \frac{2M}{n} \right)^{2/n}
\end{equation}

\subsection{AdS asymptotic region}

The firstly one needs to address how our regular solution, equations \eqref{SolucionRegular} and \eqref{SolucionFuncionMasa}, evolves as the asymptotic region is approached, roughly as $r \rightarrow \infty$. As expected, the regular solution, as $r\rightarrow \infty$,
\begin{equation}\label{AdSAsymptotia}
  \lim_{r\rightarrow \infty} f(r)  \approx 1 + \frac{r^2}{l^2} - \left(\frac{2 M}{r^{d-2n-1}} \right)^\frac{1}{n}
\end{equation}
defining an asymptotically locally AdS space as $r$ grows but also a merging into the vacuum solution counterpart \cite{Crisostomo:2000bb,Aros:2000ij}. As is well known, an asymptotically AdS behavior allows for a proper definition of conserved quantities at the boundary \cite{Aros:1999kt}. Along these lines, shadows have also been studied in spacetimes with this type of asymptotics \cite{Belhaj:2021rae,Hamil:2023zeb}.

\section{Shadows}

We study the spherically symmetric line element given by \eqref{SchwCoordinates}. Thus, the transversal section can be written as:
 \begin{equation}
 \displaystyle   d\Omega_{D-2}= d\theta^2_1 + \sum_{j=2}^{D-2} d\theta^2_j \left ( \prod_{k=1}^{j-1} \sin^2\theta_k  \right) 
\end{equation} 
We can notice that the metric tensor is independent of the angular coordinate $\theta_{D-2} = \phi$ and the temporal coordinate. In this way, we can identify the following Killing vectors:
\begin{align}
    &K_\mu= \left (-f(r),0,0,0,... \right ) \\
    & K_\mu= \left (0,0,0,...,r^2\left ( \prod_{k=1}^{D-3} \sin^2\theta_k  \right)  \right )
\end{align}
These Killing vectors lead to conserved quantities such that:
\begin{equation}
    K_\mu \dot{x}^\mu=\mbox{constant}
\end{equation}
 where the dot indicates the derivative respect to the affine parameter. We consider a movement of a  photon in an equatorial plane described by $\theta_1 = \theta_2 = \ldots = \theta_{D-3} = \pi/2$ and $\theta_{D-3}$ constant. Thus
\begin{align}
    &f(r) \dot{t}=E \\
    &r^2 \dot{\phi}=L
\end{align}
 where, in our case, $E$ and $L$ will represent the energy and angular momentum of the photon as measured by an observer at infinity.

 It is straightforward to check that the norm of the tangent vector to the geodesic is also conserved. In this line, we have:
 \begin{equation}
     \epsilon=-g_{\mu \nu} \dot{x}^\mu \dot{x}^\nu
 \end{equation}
Use $\epsilon = 0$ for null geodesics. For time-like geodesics (massive particles), $\epsilon = 1$, and for space-like geodesics, $\epsilon = -1$.  As indicated in \cite{Paithankar:2023ofw}, due to the spherical symmetry, it is sufficient to consider the motion of null geodesics in an equatorial plane. From the previous equations, we obtain:
\begin{equation}
    \frac{\dot{r}^2}{L^2} + V_{eff}(r)=\frac{1}{b^2}
\end{equation}
where the effective potential is:
\begin{equation}
    V_{eff}(r)=\frac{f(r)}{r^2}
\end{equation}
and where the impact parameter $b$ is the perpendicular distance between the position of a photon moving towards the black hole and its center.

We say that the radius of the circular orbits is the distance between the center of the black hole and a point in spacetime where photons follow a circular motion. The geodesic for these distances is a closed circle. This forms the so-called photon sphere, whose radius we will denote as $r_{sp}$.

\begin{figure}[h]
  \begin{center}
      \includegraphics[width=4in]{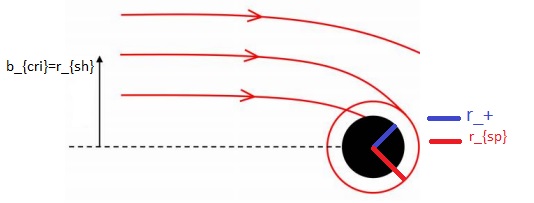}
  \caption{Scheme: We observe that photons launched with an impact parameter smaller than the shadow radius fall into the black hole. Photons with an impact parameter equal to the shadow radius fall into an unstable circular orbit. Photons launched with an impact parameter greater than the shadow radius escape toward the boundary of spacetime.} \label{FigEsquema}
  \end{center}
\end{figure}

Minima in the effective potential function correspond to stable circular orbits. Maxima, on the other hand, correspond to unstable circular orbits. It is important to mention that, in several solutions such as the Schwarzschild solution or the vacuum black hole solution of Pure Lovelock, the photon sphere is unstable because it corresponds to a maximum of the effective potential \cite{Paithankar:2023ofw}. However, it plays an important physical role. The photon sphere is the union of all closed null geodesics. Thus, there are no other types of closed null geodesics in the spacetime. Photons tangential to the photon sphere will remain on the photon sphere. However, since the photon sphere is unstable, small perturbations will cause the photon to either fall into the black hole or escape from it. In this way, the radius of the photon sphere is determined by the following equation:
\begin{equation} \label{EqMovimientoSombra}
    0= \frac{dV_{eff}}{dr} \bigg |_{r=r_{sp}} = r_{sp}^{-3} \left ( r_{sp} f'(r) |_{r=r_{sp}}-2 f(r_{sp})  \right) \Rightarrow r_{sp} f'(r) |_{r=r_{sp}}-2 f(r_{sp})=0
\end{equation}
wherethe prime indicates differentiation with respect to the radial coordinate.  Following reference \cite{Paithankar:2023ofw}, for simplicity, we assume that all light sources are placed uniformly at infinity. The critical impact parameter of the system $b_{cr}$ is the impact parameter for which the photon will precisely fall into the circular orbit around the black hole. For an impact parameter smaller than the critical impact parameter, the photon will always end up inside the black hole. For an impact parameter larger than the critical impact parameter, the photon will end up at an infinite distance from the black hole. Following \cite{Paithankar:2023ofw}, an observer located at infinity leads to the critical impact parameter being equal to the shadow radius, $r_{sh} = b_{cr}$. See the scheme in Figure \ref{FigEsquema}.  This can be computed by setting $\dot{r} = 0, r = r_{sp}$ in equation \eqref{EqMovimientoSombra}:
\begin{equation} \label{ParametroCritico}
    r_{sh}=b_{cr}=\frac{1}{\sqrt{V_{eff}}}= \frac{2 \sqrt{f(r_{sp})}}{f'(r)|_{r=r_{sp}}}
\end{equation}

As previously mentioned, the study of black holes in Lovelock gravity has attracted significant attention in recent years from a theoretical perspective. This naturally motivates the investigation of their shadow properties. In this regard, given the lack of analytical expressions for the event horizon, the photon sphere radius, and the shadow size in our case study, it becomes particularly interesting to establish numerical relationships between these shadow-related features within the context of Lovelock theories. In what follows, we propose a numerical methodology to achieve this. In the subsection below, we propose a method to achieve this. We will use the case $n=2, d=7$ as an example, but it is straightforward to check that our proposal is applicable to other values of $n$ and $d$.

\subsection{A recipe for determining the size of the photon sphere and the shadow.}

\begin{enumerate}
    \item First, since it is not possible to obtain an analytical value for the photon sphere radius from equation \eqref{EqMovimientoSombra}, we obtain the mass parameter from this equation for our RBH solution, equations \eqref{SolucionRegular} and \eqref{FuncionMasaRegular}.
    \begin{align} \label{EcuacionM(sp)}
       & M(r_{sp})= \nonumber \\
       &\frac{2^{1 - \frac{1}{n}} e^{\left( \frac{r_{sp}}{a} \right)^{\frac{d - 1}{n}}} n^2 \left( (-1 + n)! - \Gamma\left[ n, \left( \frac{r_{sp}}{a} \right)^{\frac{d - 1}{n}} \right] \right) r_{sp}^{(1 - d)} \left( (-1 + n)! - \Gamma\left[ n, \left( \frac{r_{sp}}{a} \right)^{\frac{d - 1}{n}} \right] \right)^{-\frac{1}{n}}}
{(-1 + d) r_{sp}^2 \left( -\left( \left( \frac{r_{sp}}{a} \right)^{\frac{d - 1}{n}} \right)^n + e^{\left( \frac{r_{sp}}{a} \right)^{\frac{d - 1}{n}}} n (-1 + n)! - e^{\left( \frac{r_{sp}}{a} \right)^{\frac{d - 1}{n}}} n \Gamma\left[ n, \left( \frac{r_{sp}}{a} \right)^{\frac{d - 1}{n}} \right] \right)}
    \end{align}
    
    As an example, we will study the case $n=2, d=7$ in figure \ref{FigM(sp)}. The values of the ascending curve (In the example, as we can see in the orange dashed vertical line, values greater than $\sim 6.5$ on the horizontal axis) represent the candidate values of the photon sphere radius. We use the word "candidate" because later we need to ensure that the condition $r_{sp} > r_+$ is satisfied. The vertical axis shows the values of the mass parameter. Thus, each ordered pair $\left( r_{sp}, M(r_{sp}) \right)$ represents values of the candidate photon sphere and the mass parameter in the parameter space, such that equation \eqref{EqMovimientoSombra} is satisfied, i.e., where the potential reaches an unstable maximum. On the other hand, the orange dashed horizontal line in figure \ref{FigM(sp)} represents the maximum value of the mass parameter for the parameters used. Below, we will detail what the black horizontal line in the figure represents.

\begin{figure}[h]
  \begin{center}
      \includegraphics[width=4in]{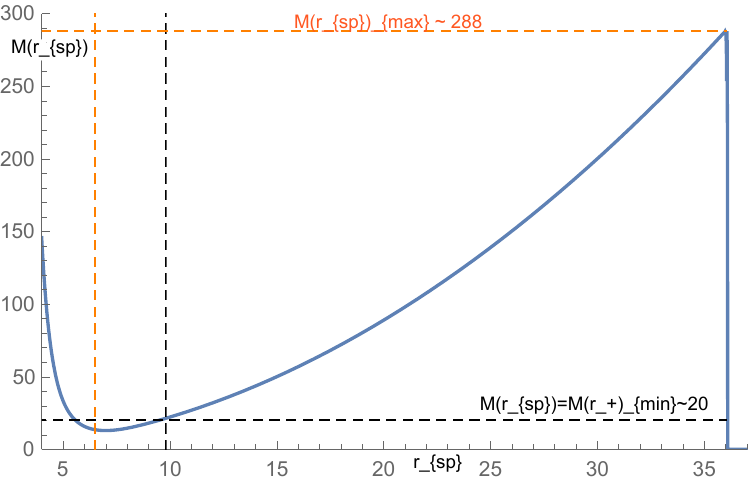}
  \caption{This figure displays the range of possible values for the photon sphere. In the horizontal axis the candidate values of the photon sphere radius $r_{sp}$. In the vertical axis the mass parameter $M(r_{sp})$. Using $n=2, d=7,a=4$}\label{FigM(sp)}
  \end{center}
\end{figure}

\item From equation \eqref{ParametroCritico} we can obtain the following relation between the photon sphere radius and the shadow's radius:
\begin{equation} \label{EcuacionSombraFoton}
  r_{sh} = \frac{r_{sp}}{\sqrt{1 + \frac{r_{sp}^2}{l^2} + \frac{2 e^{\left( \frac{r_{sp}}{a} \right)^{\frac{d - 1}{n}}} n^2    \left( (-1 + n)! - \Gamma\left[n, \left( \frac{r_{sp}}{a} \right)^{\frac{d - 1}{n}} \right] \right)}{(-1 + d) \left( \left( \frac{r_{sp}}{a} \right)^{(d-1)}  - e^{\left( \frac{r_{sp}}{a} \right)^{\frac{d - 1}{n}}} n (-1 + n)! + e^{\left( \frac{r_{sp}}{a} \right)^{\frac{d - 1}{n}}} n \Gamma\left[n, \left( \frac{r_{sp}}{a} \right)^{\frac{d - 1}{n}} \right] \right)} }}
\end{equation}

We are interested in testing the parameter values that satisfy the condition that the shadow radius is greater than the photon sphere radius, $r_{sh} > r_{sp}$. Due to the highly nonlinear nature of our equations, obtaining analytical relations for this condition becomes extremely difficult. 

From equation \eqref{EcuacionM(sp)} and figure \ref{FigM(sp)}, we extract the ordered pairs $\left( r_{sp}, M(r_{sp}) \right)$, which allow us to identify the possible values of the photon sphere radius. On the other hand, from equation \eqref{EcuacionSombraFoton}, we obtain the corresponding shadow radius values. Using this information, we determine the potential values of both the photon sphere radius and the shadow radius as given by equation \eqref{EcuacionSombraFoton}.

However, as previously mentioned, we are interested in identifying the values of the parameter $l$ that satisfy the condition that the shadow radius is greater than the photon sphere radius. To this end, using equations \eqref{EcuacionM(sp)} and \eqref{EcuacionSombraFoton}, we employ the RegionPlot command in Mathematica. This is facilitated by the fact that equation \eqref{EcuacionM(sp)} is independent of the parameter $l$.

As shown in figure \ref{FigParametroL}, for the parameters considered, the condition $r_{sch} > r_{sp}$ is fulfilled for approximately $l > 41$ across all values of $r_{sp}$. Therefore, in the following step, we must use a value of $l$ that satisfies this condition.

\begin{figure}[h]
  \begin{center}
      \includegraphics[width=4in]{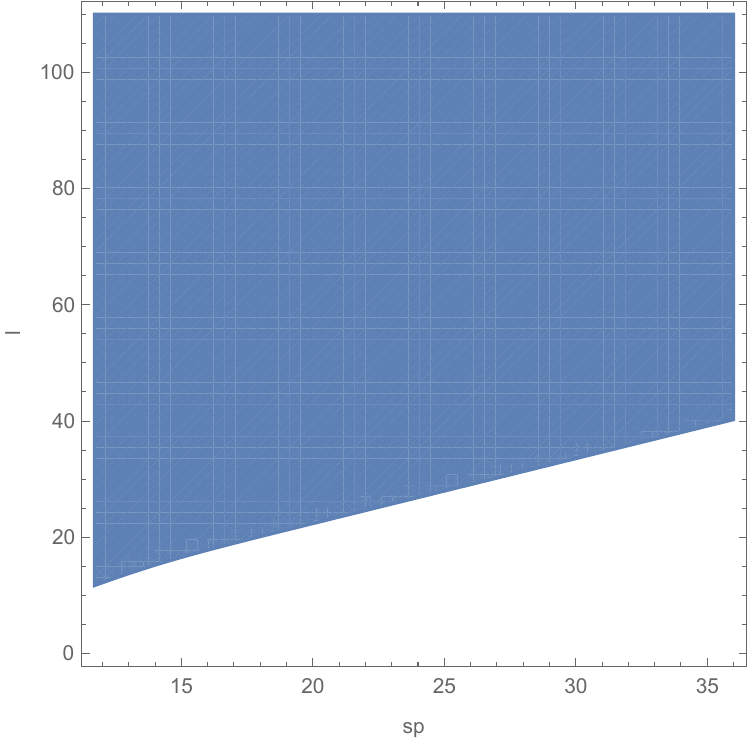}
  \caption{In blue, we observe the values of the photon sphere radius $r_{\text{sp}}$ and the AdS radius $l$ that satisfy the condition $r_{\text{sh}} > r_{\text{sp}}$}, using $n=2, d=7,a=4$ \label{FigParametroL}
  \end{center}
\end{figure}
\item To display the numerical behavior of the mass parameter, we use Equation \eqref{EqParametroMasaAgosto}, where, as previously mentioned, the values on the horizontal axis in the ascending curve correspond to the values taken by the event horizon, $\bar{r} = r_+$. 

In figure \ref{FigM(r+)}, we display the behavior of the mass parameter. We must be consistent with the previous point and choose a value of $l$ that satisfies the description provided there. In this way, on the ascending curve, we have an ordered pair of the form $\left (r_+, M(r_+) \right )$. The black dashed horizontal line represents the extremal minimum value of the mass for the chosen parameters, emphasizing that $l$ is not arbitrary.
\begin{figure}[h]
  \begin{center}
      \includegraphics[width=4in]{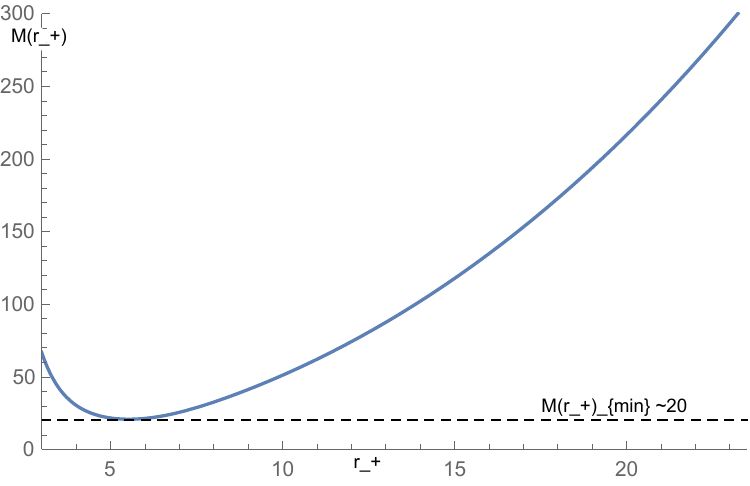}
  \caption{This figure displays the values of the event horizon corresponding to different values of the mass parameter. The values on the horizontal axis of the ascending curve represent the event horizon radius $r_+$. On the vertical axis, we have the mass parameter $M(r_+)$. Using $n=2, d=7, a=4, l=100$.}\label{FigM(r+)}
  \end{center}
\end{figure}
\item In Figure \ref{FigM(sp)}, we have ordered pairs $\left (r_{sp}, M(r_{sp}) \right )$, while in Figure \ref{FigM(r+)}, we have the ordered pairs $\left (r_+, M(r_+) \right )$. In this regard, we note the following:
\begin{itemize}
    \item The mass parameter must be the same in both cases when the remaining parameters are fixed. This is because $M$ represents the mass parameter of the solution and must have the same value for both $r_{sp}$ and $r_+$. Thus, to numerically test the relationship between $r_+$ and $r_{sp}$, we must consider that
    \begin{equation} \label{CondicionMasas}
        M(r_{sp}) = M(r_+)
    \end{equation}
    \item Following this condition, the minimum value of the mass parameter is given by the extremal value, represented by the black dashed horizontal line in Figure \ref{FigM(r+)}, $M(r_+)_{min} = M(r_{ext})$ (in the example, $\sim 20$). Therefore, the lower bound for the photon sphere radius is given by $r_{sp}$ such that $M(r_{sp}) = M(r_+)_{min} = M(r_{ext})$ (vertical line in Figure \ref{FigM(sp)}, where $r_{sp} \sim 9.8$ and $M(r_{sp} \sim 9.8) \sim 20$). 
    
    On the other hand, as shown by the orange dashed line in Figure \ref{FigM(sp)}, the parameter $M(r_{sp})$ reaches a maximum value. This last value sets the upper bound for the photon sphere and event horizon, such that condition \eqref{CondicionMasas} is satisfied ($r_{sp} \sim 36$ and $r_+ \sim 23.5$ in the examples of Figures \ref{FigM(sp)} and \ref{FigM(r+)}, respectively).
    \item Once the appropriate ranges for $r_+$ and $r_{sp}$ have been determined in the previous steps, we plot the relationship between $r_+$ and $r_{sp}$ in Figure \ref{FigHorizonFoton}. This has been done numerically, using condition \eqref{CondicionMasas} on the ordered pairs $\left (r_{sp}, M(r_{sp}) \right )$ and $\left (r_+, M(r_+) \right )$ from Figures \ref{FigM(sp)} and \ref{FigM(r+)}, respectively. As expected, we observe that the value of the photon sphere is greater than that of the event horizon. That is, photons follow a circular motion due to the geometric distortion of spacetime caused by the presence of the black hole. We also note that as $r_+$ increases, the value of $r_{sp}$ also increases in the parameter space.
\begin{figure}[h]
  \begin{center}
      \includegraphics[width=4in]{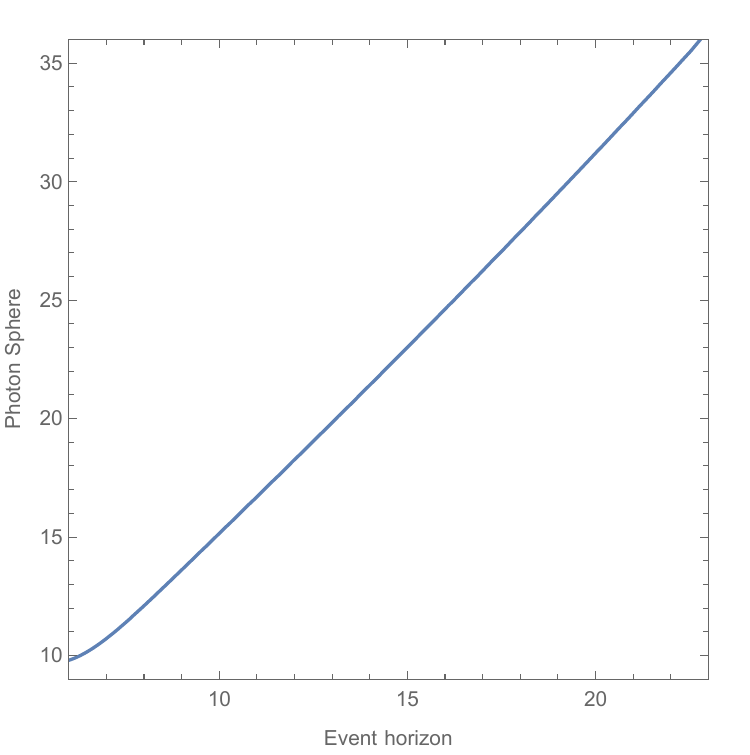}
  \caption{This figure displays the values of the photon sphere radius versus the values of the event horizon radius. The values on the horizontal axis represent the event horizon radius $r_+$. The values on the vertical axis represent the sphere photon radius $r_{sp}$. . Using $n=2, d=7, a=4, l=100$.} \label{FigHorizonFoton}
  \end{center}
\end{figure}
\end{itemize}
\item Once the appropriate ranges for $r_{sp}$ and $l$ have been determined in the previous steps, using equation \eqref{EcuacionSombraFoton}, we plot the relationship between the photon sphere radius and the shadow radius in Figure \ref{FigSpSombra}. As expected, we observe that the value of the photon sphere radius is smaller than that of the shadow radius. That is, photons that are directed with an impact parameter equal to $r_{sh}$ towards the black hole curve their trajectory towards the unstable photon sphere.
\begin{figure}[h]
  \begin{center}
      \includegraphics[width=4in]{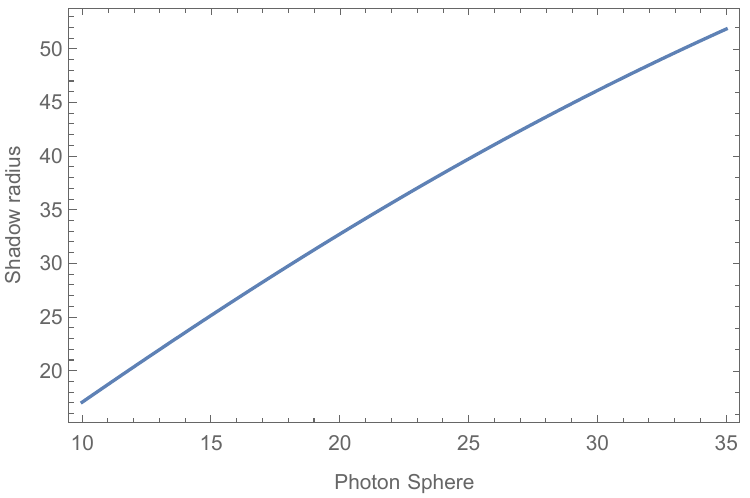}
  \caption{This figure displays the values of the shadow radius corresponding to different values of the photon sphere radius. Using $n=2, d=7, a=4, l=100$.} \label{FigSpSombra}
  \end{center}
\end{figure}
\end{enumerate}

\begin{figure}[h]
  \begin{center}
      \includegraphics[width=3in]{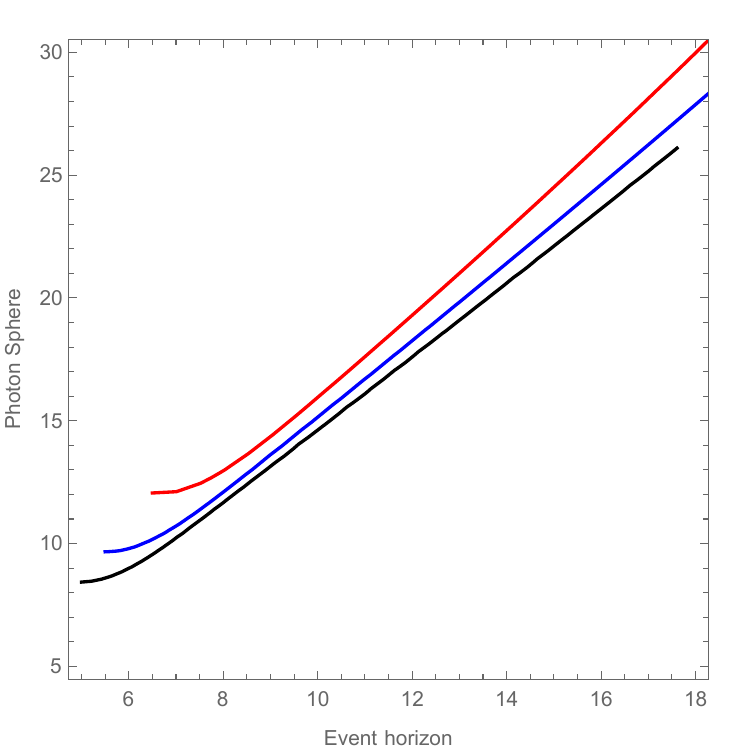}
      \includegraphics[width=3in]{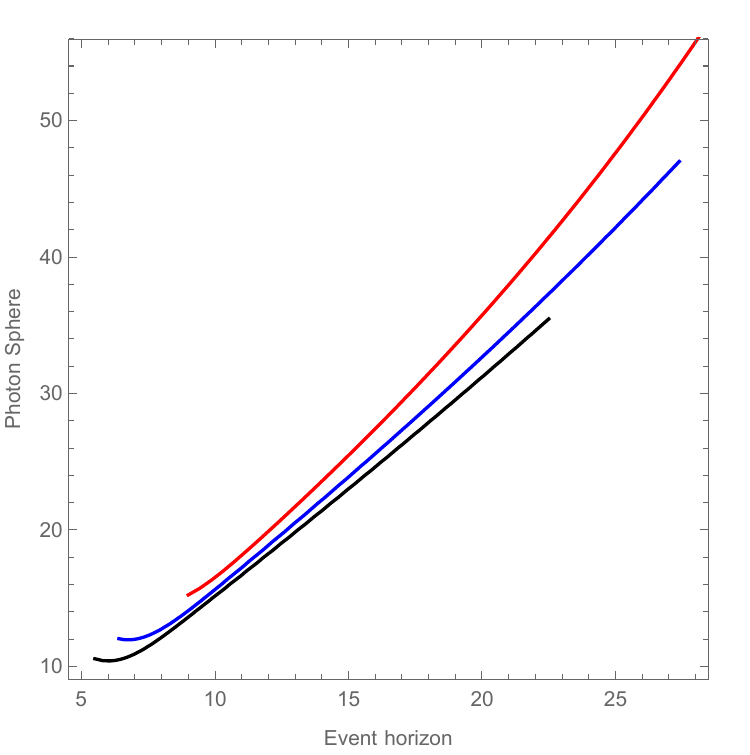}
  \caption{This figure displays the values of the photon sphere radius corresponding to different values of the event horizon radius. The values on the horizontal axis represent the event horizon radius $r_+$. The values on the vertical axis represent the sphere photon radius $r_{sp}$. In left panel we have used $n=2, a=4,l=100$, \textcolor{red}{$d=6$(red)}, \textcolor{blue}{$d=7$(blue)}, \textcolor{black}{$d=8$(black)}. In right panel we have used $n=3, a=4,l=100$, \textcolor{red}{$d=8$(red)}, \textcolor{blue}{$d=9$(blue)}, \textcolor{black}{$d=10$(black)}}\label{FigHorizonFotonN23}
  \end{center}
\end{figure}

Following the methodology described earlier, in Figure \ref{FigHorizonFotonN23} we have displayed the behavior of the event horizon versus the photon sphere radius for $n=2, d=6,7,8$ (left panel) and $n=3, d=8,9,10$ (right panel). It is important to note that, in agreement with the analysis carried out above, for each case, depending on the values of $n, d, l, a$ in the parameter space, there are different ranges for the values of the event horizon and the photon sphere radius. We can note that, if we use a common event horizon value as a reference for the same value of $n$, the photon sphere radius decreases as the number of dimensions increases.

\begin{figure}[h]
  \begin{center}
      \includegraphics[width=3in]{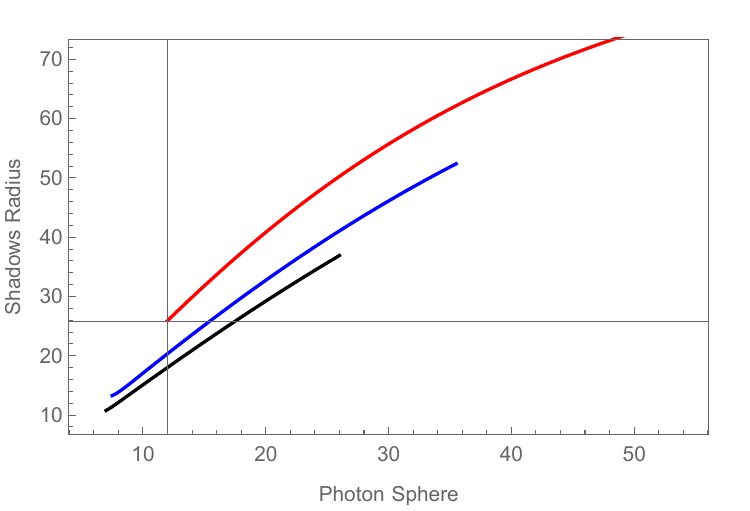}
      \includegraphics[width=3in]{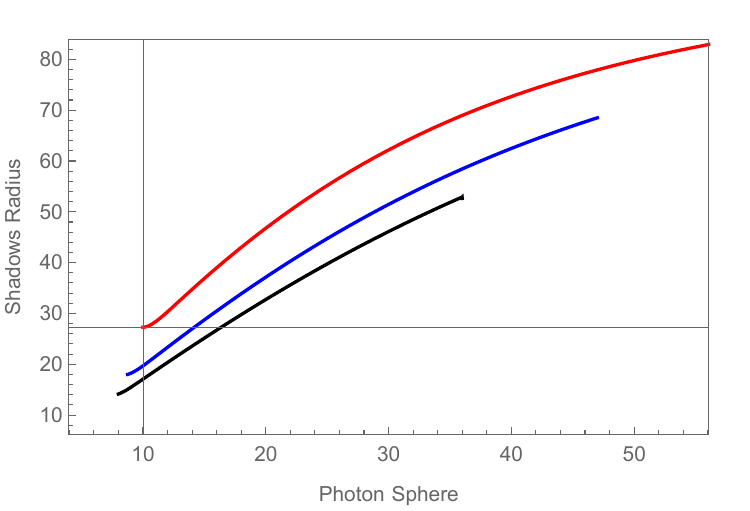}
  \caption{This figure displays the values of the shadow radius corresponding to different values of the photon sphere radius. The values on the horizontal axis represent the sphere photon radius $r_{sp}$. The values on the vertical axis represent the shadow radius $r_{sh}$. In left panel we have used $n=2, a=4,l=100$, \textcolor{red}{$d=6$(red)}, \textcolor{blue}{$d=7$(blue)}, \textcolor{black}{$d=8$(black)}. In right panel we have used $n=3, a=4,l=100$, \textcolor{red}{$d=8$(red)}, \textcolor{blue}{$d=9$(blue)}, \textcolor{black}{$d=10$(black)}}\label{FigFotonSombraN23}
  \end{center}
\end{figure}

In Figure \ref{FigFotonSombraN23}, we have displayed the behavior of the photon sphere radius versus the black hole shadow radius for $n=2, d=6,7,8$ (left panel) and $n=3, d=8,9,10$ (right panel). In the same way as in the previous figure, and in agreement with the analysis carried out above, for each case, depending on the values of $n, d, l, a$ in the parameter space, there are different ranges for the values of the photon sphere radius and the shadow radius. We can note that, if we use a common photon sphere radius value as a reference for the same value of $n$, the size of the black hole shadow radius tends to decrease as the number of dimensions increases.

From a physical point of view, our methodology allows us to estimate the parameter ranges for which the condition that the shadow radius fully encloses the photon sphere radius, and the latter, in turn, fully encloses the event horizon, is satisfied , that is, the condition $r_+ < r_{\text{sp}} < r_{\text{sh}}$ holds. This is consistent with the analytical treatment of the vacuum Pure Lovelock case presented in Reference \cite{Paithankar:2023ofw}, which appears to be the only analytically solvable case currently known for Lovelock gravity.

In this way, our numerical approach could eventually be used for a deeper investigation in a future work, enabling, for example, the estimation of the deflection angle or the characterization of a linearly uniformly accelerated trajectory. In addition to the above, we would like to point out that establishing a direct comparison between the AdS radius values used in our figures and those estimated by other authors in the context of Lovelock theories would require , at the very least, theoretical justification regarding the behavior of the full set of black hole parameters within the AdS–Lovelock framework. This latter aspect goes beyond the scope of the present work. As mentioned, the values of the AdS radius currently available in the literature for Lovelock gravity remain speculative \cite{Chernicoff:2016uvq,Camanho:2014apa}.

\section{Thermodynamics}

First of all, it is worth noting that the energy can be computed as a conserved charge by following the methodology in Reference \cite{Aros:2000ij}, which involves a regularization process. In this case, the conserved charge associated with the energy, computed using a timelike Killing vector, is obtained in an analogous way to Reference \cite{Aros:2019quj}:
\begin{equation}\label{Energy}
E = M
\end{equation}
In this way, the black hole energy is directly related to the parameter $M$.

On the other hand, in Reference \cite{Estrada:2021kuj} it was shown, using the conditions $f(r_+, M) = 0$ and $\delta f(r_+, M) = 0$, which can be viewed as constraints on the evolution along the space of parameters, that the first law of thermodynamics takes the form: 
\begin{equation}
    \frac{\partial {m}}{\partial M}dM=  \left ( \frac{1}{4\pi} f'\big |_{r=r_+}  \right )  \left (2 \pi n r_+^{d-2n-1} \left (1+ \frac{r_+^2}{l^2} \right )^{n-1} dr_+\right ).
\end{equation}

The above equation can be rewritten as:
\begin{equation}
    du=TdS,
\end{equation}
where we can identify the temperature and entropy terms as:
\begin{align}
    T=&\frac{1}{4\pi} f'\big |_{r=r_+} \\
   dS=& 2 \pi n r_+^{d-2n-1} \left (1+ \frac{r_+^2}{l^2} \right )^{n-1} dr_+
\end{align}
where $du$ corresponds to a local definition of the variation of the energy at the horizon. On the other hand, the definitions of entropy and temperature coincide with those of the vacuum case. That is, our matter sources influence the behavior of the aforementioned definition $du$. 

In this section, we study the thermodynamics at the event horizon, that is, where the mass parameter satisfies $M \geq M_{\text{min}}$, where $M_{\text{min}}$, as previously mentioned, is associated with the extremal radius, where the inner and event horizons coincide, $r_{\text{ext}} = r_- = r_+$. It is worth emphasizing, as mentioned earlier, that an interesting feature of our solution is that, for a value of $M$ slightly greater than $M_{\text{min}}$, the behavior of the mass parameter in our solution becomes indistinguishable from the vacuum case. Therefore, under this condition, the event horizons of both solutions coincide. As also discussed, this implies that the geometry of both cases becomes indistinguishable beyond the event horizon.

\subsection{Temperature}

The temperature is defined by
\begin{eqnarray}
 4\pi T &=& \frac{d}{dr} f(r) \Big |_{r=r_+} = 2\frac{r_{+}}{l^2} + \frac{(d-2n-1)}{nr_{+}}\left(\gamma + \frac{r_{+}^2}{l^2}\right) \nonumber\\
 &-&  \frac{1}{n!}\frac{\exp{\left( -\left(\frac{r_+}{a}\right)^{\frac{d-1}{n}}\right)}}{\left(1 - e^{ -\left(\frac{r_+}{a}\right)^{\frac{d-1}{n}}} \displaystyle \sum^{n-1}_{i=0} \frac{1}{i!} \left(\frac{r_+}{a}\right)^{\frac{i}{n} (d-1)}\right)} \left(\frac{r_+}{a}\right)^{\frac{n-1}{n} (d-2n-1)} \label{Temperature}
\end{eqnarray}
In Eq. (\ref{Temperature}), we can notice that the first two terms are independent of $a$, and furthermore, they match the expression for the temperature found in \cite{Aros:2000ij,Crisostomo:2000bb} for the vacuum LnFDGS AdS solution. Regarding the third term in Eq. (\ref{Temperature}), we can analytically observe that for a scenario where $r_{+} \gg a$, condition
\begin{equation}\label{Convergency}
  \lim_{r_+/a \rightarrow \infty} \frac{\exp{\left( -\left(\frac{r_+}{a}\right)^{\frac{d-1}{n}}\right)}}{\left(1 - e^{ -\left(\frac{r_+}{a}\right)^{\frac{d-1}{n}}}\sum^{n-1}_{i=0} \frac{1}{i!} \left(\frac{r_+}{a}\right)^{\frac{i}{n} (d-1)}\right)} \left(\frac{r_+}{a}\right)^{\frac{n-1}{n} (d-2n-1)} \approx 0
\end{equation}
holds. Thus, we find analytically that in a regime where $a \ll r_+$, the temperature of the vacuum LnFDGS AdS solution becomes indistinguishable from that of our RBH LnFDGS AdS solution. This regime could be of physical interest in the case where $a$ is of the order of the Planck scale, while $r_+$ corresponds to a realistic radius—on the order of the radius of the rotating black hole observed in LIGO GW150914, $r_+ \sim 103,\text{km}$—which is much larger than the Planck scale.

Numerically, we can note that for the value of $M = M_{min}$ where $r_- = r_+ = r_{ext}$, the temperature vanishes. In this sense, it is well known that the zero temperature point is associated with a black remnant, known as what remains once evaporation stops. In this context, in Figure \ref{TemperatureGraf} below, we observe that a remnant indeed forms in this extremal case.

In Fig. \ref{TemperatureGraf}, the generic numerical behavior of the temperature as a function of the event horizon $r_+$ is displayed. We can observe that this behavior is generic for other values of the parameters involved. In this figure, the horizontal axis can be considered to be of the order of Planck units, $\ell_p$. Two curves are shown. The blue curve represents the temperature of the vacuum LnFDGS AdS solution \cite{Aros:2000ij,Crisostomo:2000bb} as a function of the event horizon. The orange curve corresponds to the temperature of our RBH LnFDGS AdS solution. It can be observed that there exists a finite value slightly greater than the extremal radius, $r_* > r_{ext}$, where the temperatures of both cases become indistinguishable. 

It is of interest to discuss what happens for values smaller than $r_*$, that is, radial values prior (from left to right) to the point where both temperatures become indistinguishable:

\begin{itemize}
\item The temperature of the vacuum solution increases and hypothetically diverges in the limit $r_+ \to 0$.
\item On the other hand, in the case of the regular solution, we can note that the temperature decreases until it vanishes for the value $M = M_{\text{min}}$, where $r_- = r_+ = r_{\text{ext}}$.
\end{itemize}

In this sense, it is well known that the point of zero temperature is associated with a black hole remnant, what remains once evaporation ceases. In this context, in Figure \ref{TemperatureGraf} below, we observe that a remnant indeed forms in this extremal case.

Thus, at the small Planck scale, quantum effects would emerge. This implies that, instead of the temperature diverging as in the vacuum case, the matter sources proposed in this work lead the temperature to decrease until a black hole remnant is reached at $T = 0$ and $r_+ = r_{\text{ext}}$.

In other words, the behavior of temperature suggests that our forms of matter in the energy-momentum tensor play a role in eliminating divergences in LnFDGS AdS theories. This is analogous to the role of non-commutative matter sources in General Relativity (GR), which in turn play the same role in quantum field theory and string theory \cite{Nicolini:2005vd,Smailagic:2012cu} . Therefore, it would be of interest to study the role of our LnFDGS AdS energy sources in quantum field theory in a future work. As shown, the role of our matter sources is to cool the black hole in the final stage. Consequently, analogous to the non-commutative $4D$ case studied in GR \cite{Nicolini:2005vd}, this could be interpreted as a suppression of the quantum back-reaction of Hawking radiation once the temperature has already reached its maximum, as the black hole emits progressively less energy. As mentioned, these effects could be tested at Planck scales.

\begin{figure}[h]
  \begin{center}
      \includegraphics[width=6in]{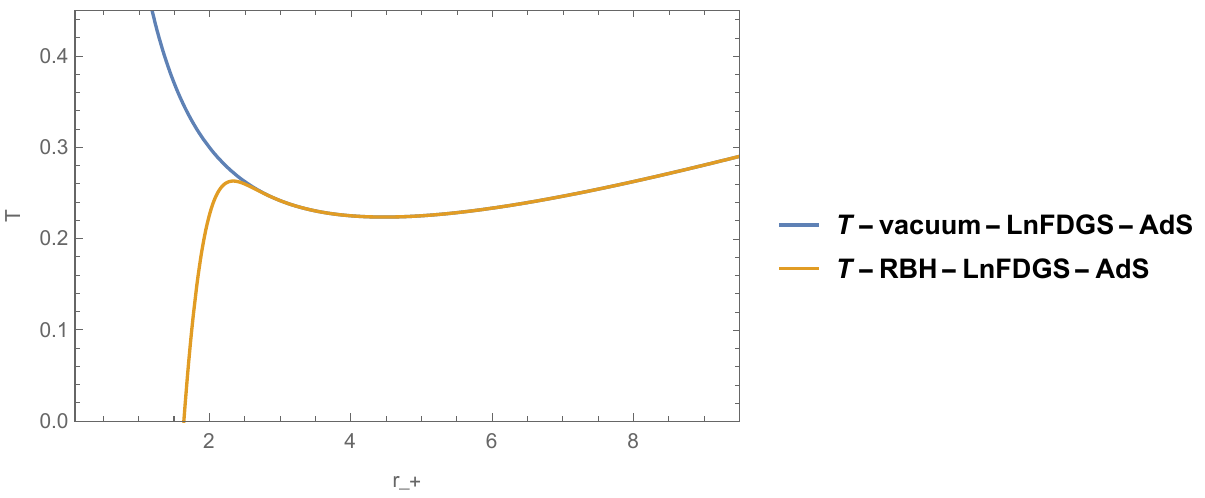}
  \caption{This figure displays the values of the temperature corresponding to different values of the event horizon radius for $d=6$, $n=2$, $a=1$ and $l=10$}\label{TemperatureGraf}
  \end{center}
\end{figure}

\subsection{Heat Capacity and radial evolution}

We use the standard definition of heat capacity:

\begin{equation}\label{HeatCapacityGen}
  C = \left ( \frac{dT}{dM}\right)^{-1}= \frac{dM}{dr_{+}} \left(\frac{dT}{dr_{+}}\right)^{-1}
\end{equation}

In this section, we are interested in identifying the regions where the heat capacity is positive (stable), negative (unstable), and where it diverges, indicating the presence of phase transitions. In this connection, from Figure \ref{TwoMasses}, the derivative $\frac{dM}{dr_+}$ is always positive, i.e., $\frac{dM}{dr_+} \geq 0$.

Since the sign of the derivative $\frac{dT}{dM} = \left(\frac{dT}{dr_+}\right) \left(\frac{dM}{dr_+}\right)^{-1}$ depends only on the sign of the derivative $\frac{dT}{dr_+}$, it follows that the sign of the heat capacity also depends solely on this latter derivative. In connection with the previous paragraph, from Figure \ref{TemperatureGraf}, we can observe that, for our solution, the temperature exhibits three regions from left to right:
i) $\frac{dT}{dr_+} > 0$ due to the influence of matter sources at small scales;
ii) and iii) which coincide with the vacuum LnFDGS AdS case, where $\frac{dT}{dr_+} < 0$ and $\frac{dT}{dr_+} > 0$, respectively. Thus, in our solution, there would be two phase transitions (analyzed from right to left):
i) $C > 0$ where $\frac{dT}{dM} > 0$; ii) $C < 0$ where $\frac{dT}{dM} < 0$ (these two coincide with the vacuum LnFDGS AdS case, where both solutions become indistinguishable); iii) $C > 0$ where $\frac{dT}{dM} > 0$.

Once both the vacuum solution and our regular black hole (RBH) become distinguishable, the heat capacities of the two solutions also differ at small scales. In this way, we can observe that: Moving from right to left, while the heat capacity of the vacuum AdS solution remains negative, implying that complete evaporation, as the parameter $M$ decreases toward $M \to 0$, would cause the temperature to diverge to infinity, in our regular AdS solution, the matter sources induce a new phase transition such that the heat capacity changes from negative to positive. This implies that, as the mass decreases to reach the extremal value $M = M_{\text{min}}$, the temperature decreases until a black hole remnant is reached at $T = 0$, at which point the evaporation process halts. As mentioned earlier, this could occur at scales close to the Planck scale. That is, the nature of our matter sources, along with the avoidance of singularity formation, would lead to the existence of a remnant.

\section{DISCUSSION AND SUMMARIZE}

As mentioned in the introduction, a model was recently proposed in Ref.~\cite{Estrada:2024uuu} to suppress physical singularities in Pure Lovelock (PL) gravity, where the energy density encodes the gravitational information of the vacuum solution through the Kretschmann scalar. However, in the PL framework, the inclusion of a negative cosmological constant is unfeasible, as its presence leads to the appearance of a curvature singularity in the spherically symmetric case.

To address the aforementioned issue, namely, the introduction of an energy density analogous to that in Ref.~\cite{Estrada:2024uuu} within Lovelock theories under the presence of a negative cosmological constant, in this work, instead of considering the PL action, which includes only a single term in the Lovelock series, we adopt the definition of the coupling constants given in Refs.~\cite{Crisostomo:2000bb,Aros:2000ij}. By including all terms in the Lovelock series up to order $n = \left[ d/2 \right]$, the resulting equations of motion possess an $n$-fold degenerate AdS ground state. In this way, the inclusion of a negative cosmological constant does not lead to the emergence of curvature singularities.

Given the structure of the equations of motion in LnFDGS theories, providing a form for the energy density analogous to the one described earlier is not straightforward. This is because several additional terms would appear when directly substituting the Kretschmann scalar of the vacuum LnFDGS solution. To address this, we have defined an appropriate gauge curvature tensor for LnFDGS, $F^{\text{AdS}}_{\mu\nu}$, which we have called the Poincaré AdS-like curvature. From this, we will provide an alternative version of the Kretschmann scalar associated with gravitational tension. In this way, using the structure provided for gravitational tension, we have defined the energy density, which maintains the physical arguments provided in reference \cite{Estrada:2023pny} for PL.

The obtained solution is such that, for a mass parameter value slightly greater than $M_{\text{min}}$ (associated with an extremal radius), our RBH solution LnFDGS becomes indistinguishable from its vacuum counterpart LnFDGS \cite{Crisostomo:2000bb,Aros:2000ij} inside the event horizon, from a radial value $r = r_- < r_+$ up to infinity. If $r_*$ is of the order of the Planck scale, the geometric differences between both solutions, particularly the suppression of the singularity, would occur at quantum scales.

On the other hand, due to the high theoretical interest that the study of black holes in Lovelock gravity has garnered, it is natural to explore the properties of their shadows. However, in most solutions within this theory—such as our case of interest, the LnFDGS AdS solution—the analytical study of these shadows becomes challenging. In this regard, related to the study of shadows, a method has been proposed to obtain numerical graphical relationships between the event horizon, the photon sphere radius, and the shadow size in our case study. The case $n=2$, $d=7$ has been used as an example to explain our procedure. However, it is straightforward to verify that our proposal is applicable to other values of $n$ and $d$. We have also presented the graphical behavior for the cases $n=2$, $d=6,7,8$ and $n=3$, $d=8,9,10$. We note that, when using a common event horizon value as a reference for the same value of $n$, the photon sphere radius decreases as the number of dimensions increases. Additionally, when using a common photon sphere radius value as a reference for the same value of $n$, the size of the black hole shadow radius tends to decrease as the number of dimensions increases. It is also important to mention that, in all the cases studied numerically, the condition $r_+ < r_{\text{sp}} < r_{\text{sh}}$ is satisfied.

From the analysis of the temperature, it can be observed that there exists a finite value slightly greater than the extremal radius, $r_* > r_{ext}$, where the temperatures of both cases also become indistinguishable. Thus, at the Planck scales, quantum effects would arise, meaning that instead of the temperature evolving to infinity as in the vacuum LnFDGS case, the matter sources proposed in this work cause the temperature to decrease until a black hole remnant is reached at $T = 0$ and $r_+ = r_{ext}$.

Regarding radial evolution and heat capacity, once the heat capacity of both the vacuum solution and our regular solution is distinguible at short scales, we observe a key difference: while the heat capacity of the vacuum AdS solution remains negative, implying that complete evaporation, as the parameter $M$ decreases toward $M \to 0$, would cause the temperature to diverge to infinity. In our regular AdS solution, the matter sources induce a new phase transition such that the heat capacity changes from negative to positive.
This implies that as the mass decreases and approaches the extremal value $M = M_{\text{min}}$, the temperature also decreases until a black hole remnant is reached at $T = 0$, at which point the evaporation process halts. As mentioned earlier, this could occur at scales close to the Planck scale. That is, the nature of our matter sources, along with the avoidance of singularity formation, would lead to the existence of a remnant.

\section*{Acknowledgements}

This work of RA was partially funded through FONDECYT-Chile 1220335. Milko Estrada is funded by the FONDECYT Iniciaci\'on Grant 11230247.

\bibliography{2025mybib}

\end{document}